\documentclass[aps,pre,twocolumn,superscriptaddress,showpacs,showkeys ]{revtex4-1}

\usepackage{graphicx}                   
\usepackage{hyperref}                   
\usepackage{amsmath,amssymb}            
\usepackage{epstopdf}
\usepackage{subcaption}					

\usepackage{color}
\definecolor{orange}{rgb}{1,0.5,0}
\definecolor{brown}{rgb}{0.65, 0.16, 0.16}
\definecolor{phlox}{rgb}{0.87, 0.0, 1.0}

\graphicspath{{figs/}}					

\begin{document}

\title{The elastic backbone phase transition in the Ising model}

\author{M. N. Najafi}
\affiliation{Department of Physics, University of Mohaghegh Ardabili, P.O. Box 179, Ardabil, Iran}
\affiliation{Computational Physics, IfB, ETH Zurich, Stefano-Franscini-Platz 3, CH-8093 Zurich, Switzerland}
\email{morteza.nattagh@gmail.com}

\author{J. Cheraghalizadeh}
\affiliation{Department of Physics, University of Mohaghegh Ardabili, P.O. Box 179, Ardabil, Iran}
\email{jafarcheraghalizadeh@gmail.com}

\author{H. J. Herrmann}
\affiliation{Computational Physics, IfB, ETH Zurich, Stefano-Franscini-Platz 3, CH-8093 Zurich, Switzerland}
\email{morteza.nattagh@gmail.com}
\affiliation{Departamento de Fsica, Universidade Federal do Ceara, 60451-970 Fortaleza(Brazil)}
\affiliation{ESPCI, CNRS UMR 7636 - Laboratoire PMMH, 75005 Paris (France)}

\begin{abstract}
The two-dimensional (zero magnetic field) Ising model is known to undergo a second order para-ferromagnetic phase transition, which is accompanied by a correlated percolation transition for the Fortuin-Kasteleyn (FK) clusters. In this paper we uncover that there exists also a second temperature $T_{\text{eb}}<T_c$ at which the elastic backbone of FK clusters undergoes a second order phase transition to a dense phase. The corresponding universality class, which is characterized by determining various percolation exponents, is shown to be completely different from directed percolation, proposing a new anisotropic universality class with $\beta=0.54\pm 0.02$, $\nu_{||}=1.86\pm 0.01$, $\nu_{\perp}=1.21\pm 0.04$ and $d_f=1.53\pm 0.03$. All tested hyper-scaling relations are shown to be valid.
\end{abstract}


\maketitle

\section{Introduction}

The geometrical approach to thermal systems has proved to be very fruitful, especially in the vicinity of critical points. The effectiveness of the correspondence between local and global properties has led to the study of various geometrical quantities in thermal systems, like the $q$-state Potts model~\cite{potts1952some,janke2004geometrical}, the two-dimensional electron gas~\cite{najafi2018percolation}, the spin glass~\cite{bernard2007possible} and the modified Ising models~\cite{najafi2016universality,najafi2016monte}. Backbone and elastic backbone (EB, the set of shortest paths) of the geometrical and the Fortuin-Kasteleyn (FK) clusters are examples of such extended objects, whose fractal structure can be found in optimal paths~\cite{herrmann1988fractal} and interfaces~\cite{Cardy2005Sle}, which can be processed via Schramm-Loewner evolution (SLE)~\cite{najafi2015observation}. Actually the criticality of the original model induces fractality of these extended objects. More precisely when the thermal model experiences a second order phase transition, it can be equivalently described as a percolation transition of FK clusters, which are fractal~\cite{vasseur2012critical}. The elastic backbone (EB) will serve here as a geometrical object that can be employed to lighten some aspects of geometrical and also FK clusters.\\

The EB in disordered systems is the subset of the backbone that would give the first contribution to a restoring force, when the system is elongated. The EB determines the resistance of the system under tension, whose characterization involves the determination of its fractal dimension, optimal path traces, etc.~\cite{herrmann1988fractal}. A new type of transition in classical percolation for the EB was discovered in Ref.~\cite{sampaio2018elastic}. It was observed that the EBs of the percolation model on the tilted square lattice and also on the triangular lattices undergo a second order phase transition at some $p_{\text{eb}}>p_c$, above which the EBs become dense. Various new exponents were calculated. Shortly thereafter it was shown that the set of the shortest paths in ordinary percolation system behaves just like the backbone of directed percolation (DP)~\cite{deng2018elastic}. A question rises here whether such a transition is also seen in thermal systems, e.g. the Ising model as the simplest one.\\
The fact that many binary systems can be mapped to the Ising model, makes such a study worthy. Examples are the oxygen configuration in YBCO planes~\cite{najafi2016universality,pekalski1994monte}, protein folding~\cite{munoz2001can}, position configuration of metallic nano-particles in random media~\cite{cheraghalizadeh2018gaussian}, the position of non-permeable rocks in reservoirs~\cite{cheraghalizadeh2017mapping}, etc. It may be seen as a way of making a percolation system correlated~\cite{delfino2009field}.\\
This paper is devoted to investigate the geometrical properties of the EBs of the FK clusters of the Ising model in terms of temperature. To this end we define the Ising model on the tilted square lattice and extract its various critical exponents. Interestingly we observe a threshold temperature $T_{\text{eb}}<T_c$ below which the EBs become dense. We show that all tested hyperscaling relations hold, and the anisotropic universality class is clearly different from the DP universality class. \\

The paper has been organized as follows: In the next section we shortly introduce the FK representation of the $q$-state Potts model. Section~\ref{res} has been devoted to the numerical details and results. We close the paper by a conclusion.

\section{The Fortuin-Kasteleyn (FK) representation of the Ising model}\label{FKCluster}
\label{sec:model}
 The FK formulation provides a geometrical description of the $q$-state Potts model. The determination of these geometrical properties is of especial importance in the context of critical phenomena. The FK clusters of the $q$-state Potts model describe the critical behavior. The $q$-state Potts model is defined by the following Hamiltonian:
 \begin{equation}
 H=-K\sum_{\left\langle i,j\right\rangle}\left( \delta_{\sigma_i,\sigma_j}-1\right) -h'\sum_{i}\delta_{\sigma_i,1}, \ \ \ \ \ \sigma_i=1,2,...,q
 \label{Eq:Potts}
 \end{equation}
 where $K$ is the coupling constant, $\sigma_i$ and $\sigma_j$ are the spins at the sites $i$ and $j$ respectively (taking $q$ states), $h'$ is the magnetic field, and $\left\langle i,j \right\rangle$ shows that the sites $i$ and $j$ are nearest neighbors. The celebrated FK representation of $q-$state Potts model is expressed via the following partition function (for the zero magnetic field):
 \begin{equation}
 Z_{FK}=\sum_{\Gamma}p^b(1-p)^{B-b}q^{N_c}
 \end{equation}
 in which $p=1-e^{-K}$, $N_c$ is the number of clusters, and $\left\lbrace \Gamma\right\rbrace $ denotes the set of bond configurations specified by $b$ occupied bonds and $\bar{b}\equiv B-b$ broken bonds, where $B$ is the total number of bonds in the configuration $\Gamma$. For $q\leq 4$, where the $q$-state Potts model undergoes a continuous phase transition, these clusters percolate at the critical temperature. At the technical level, the FK clusters are also useful to reduce the critical slowing down, which is known as the Swendsen-Wang algorithm~\cite{swendsen1987nonuniversal}. In this approach FK clusters are used as the objects to be updated at each Monte Carlo step. If we take $\tau$ and $\sigma$ as two independent exponents, defined by $P(n)\sim n^{-\tau}\exp\left[-\theta n \right] $ in which $P(n)$ is the cluster distribution giving the average number density of clusters of $n$ sites and $\theta\sim (T-T_c)^{1/\sigma}$, then the standard geometrical exponents of the percolation theory are given by:
 \begin{equation}
 \begin{split}
 &\alpha=2-\frac{\tau-1}{\sigma}, \ \ \beta=\frac{\tau-2}{\sigma}, \ \ \gamma=\frac{3-\tau}{\sigma} \\
 &\eta=2+d\frac{\tau-3}{\tau-1}, \ \ \nu=\frac{\tau-1}{d\sigma}, \ \ d_f=\frac{d}{\tau-1} 
 \end{split}
 \label{Eq:hyperscaling}
 \end{equation}\\
 in which $\alpha$ is the exponent of the density of clusters (determined by the divergence of its third derivative with respect to temperature), $\beta$ is the exponent of the number density of the percolating cluster, $\gamma$ is the exponent of density fluctuations, $\eta$ is the Fisher exponent (anomalous dimension in the Green function), $\nu$ is the exponent of correlation length, and $d_f$ is the cluster fractal dimension. Therefore some hyper-scaling relations relate these exponents, the most important ones being $\alpha=2-\nu d$, $d_f=\frac{1}{2}(d+2-\eta)=d-\beta/\nu$, $2\beta+\gamma=d\nu$. The latter hyper-scaling relation is violated for the EB transition of the percolation model~\cite{sampaio2018elastic}. \\
 
 The Ising model is given ($q=2$ Potts model) by (up to an additive constant):
 \begin{equation}
 H=-J\sum_{\left\langle i,j\right\rangle}s_is_j -h\sum_{i}s_i, \ \ \ \ \ s_i=\pm 1
 \label{Eq:Ising}
 \end{equation}
 in which $J=\frac{1}{2}K$, and $h=\frac{1}{2}h'$. $J>0$ corresponds to positively correlated nearest neighbors whereas $J<0$ is for negatively correlated ones. The temperature $T$ controls the disorder in the system. The FK clusters are simply obtained by bond-diluting the geometric spin cluster, i.e. the connected cluster comprised by the same spins. In this bond-dilution, one removes the bonds between nearest neighbors with the probability $p=1-e^{-2J}$.\\

For $h=0$ the model is well-known to exhibit a non-zero spontaneous magnetization per site $M=\lim_{h\to‎ 0}\left\langle \sigma_i\right\rangle $ at temperatures below the critical temperature $T_c$. In fact there are two transitions in the Ising model: the magnetic (paramagnetic to ferromagnetic) transition (mentioned above) and the percolation transition (in which the FK cluster percolate and become fractal). For the 2D regular Ising model at $h=0$ these two transitions occur simultaneously~\cite{delfino2009field}, although it is not the case for all versions of the Ising model, e.g. for the site-diluted Ising model~\cite{najafi2016monte}.\\
\section{Results}\label{res}
\begin{figure*}
	\begin{subfigure}{0.30\textwidth}\includegraphics[width=\textwidth]{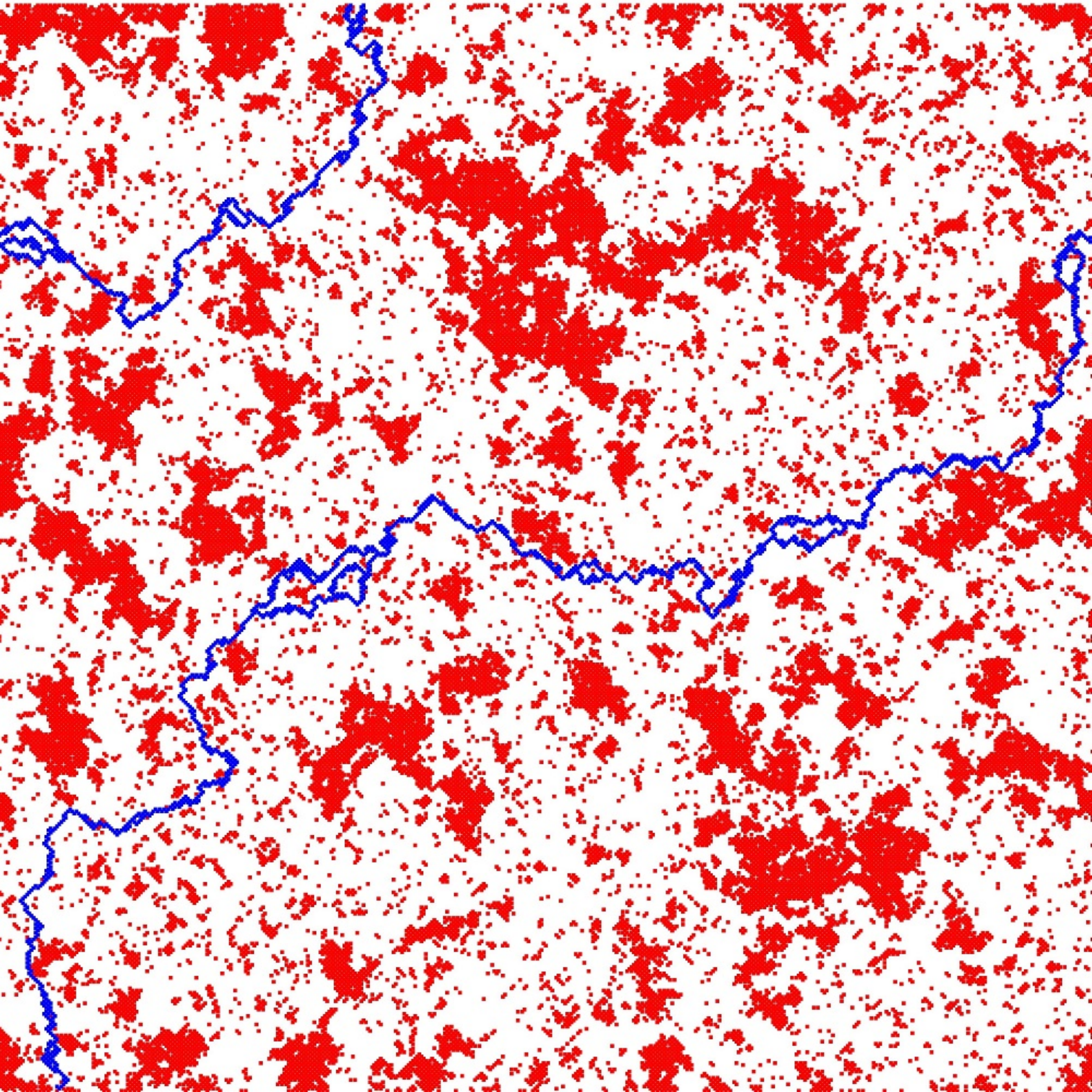}
		\caption{}
		\label{fig:HighT}
	\end{subfigure}
	\begin{subfigure}{0.30\textwidth}\includegraphics[width=\textwidth]{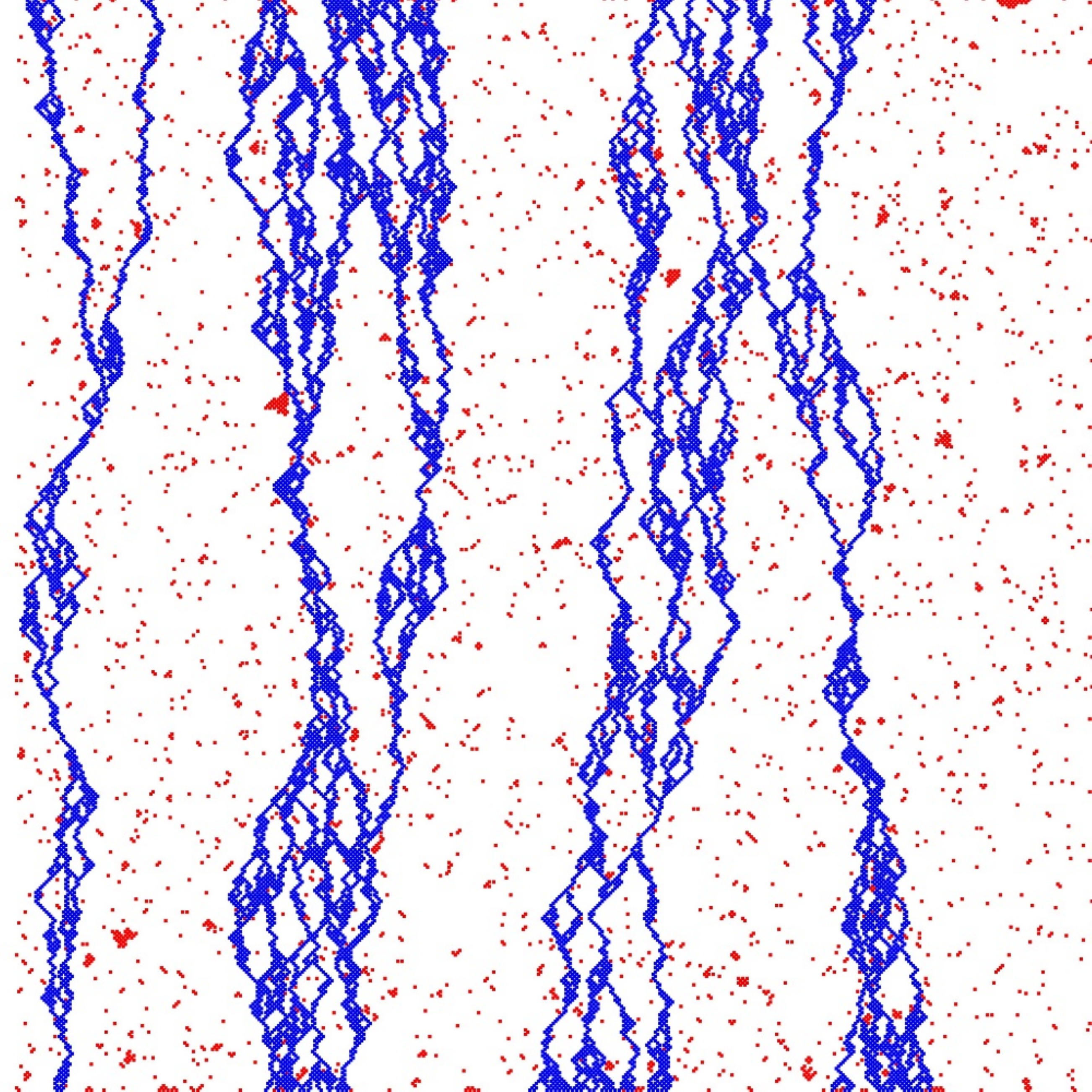}
		\caption{}
		\label{fig:T_eb}
	\end{subfigure}
	\centering
	\begin{subfigure}{0.30\textwidth}\includegraphics[width=\textwidth]{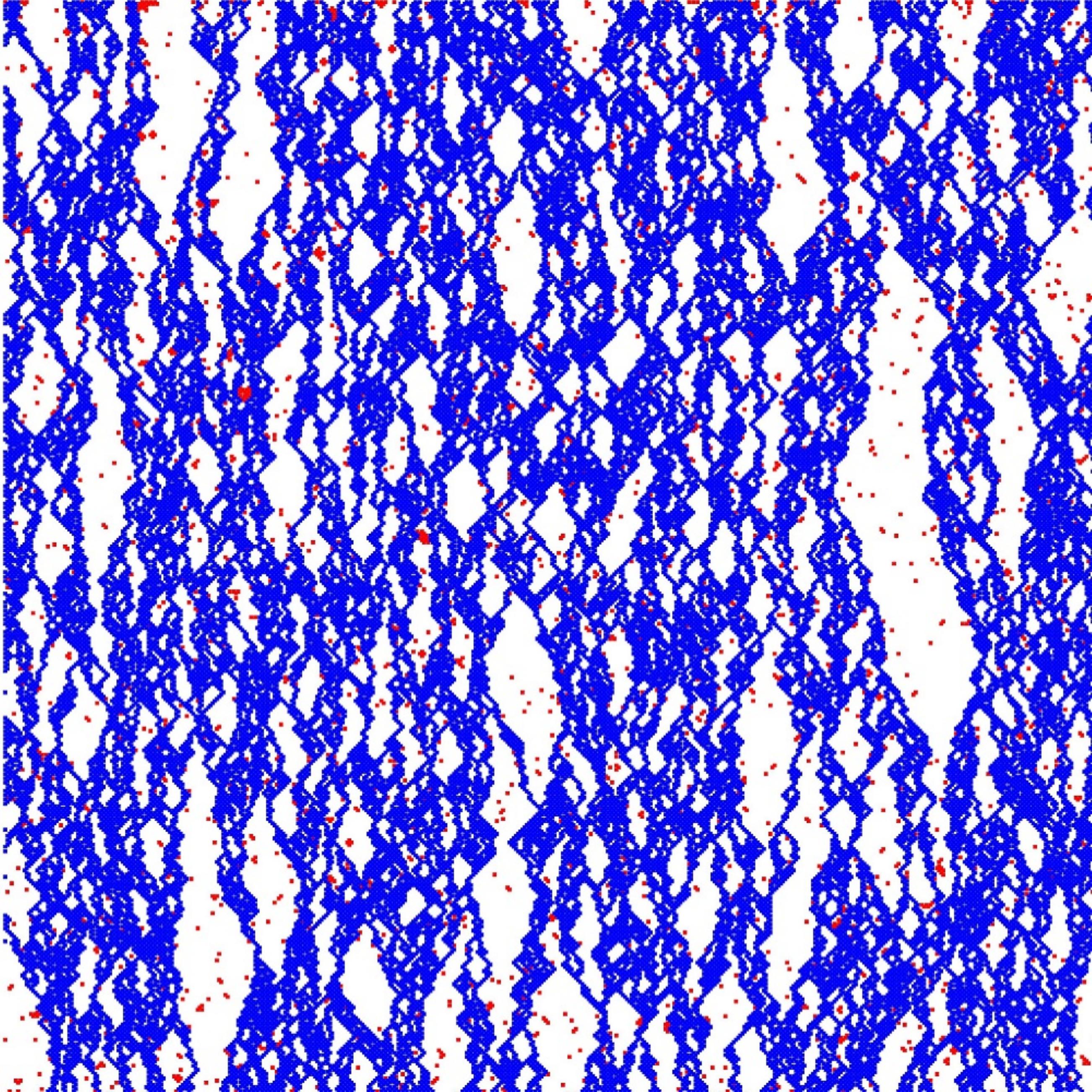}
		\caption{}
		\label{fig:LowT}
	\end{subfigure}
	\caption{(Color online):Images of the EB for (a) $T=T_c>T_{\text{eb}}$ (dilute phase) (b) $T=T_{\text{eb}}$ (critical value) and (c) $T=1.7<T_{\text{eb}}$ (dense phase) on the tilted square lattice with $L=256$. The white/red sites are majority/minority spins that do not belong to the EBs, and the blue sites belong to the EBs (FK clusters are not shown). }
	\label{Fig:IsingSamples}
\end{figure*}
As a spanning object, the elastic backbone (EB) is a geometrical subset of the spanning cluster that contains important information about the geometry of the cluster, since it is the set of points which react first to an external tension. It defines a new type of transition in ordinary percolation~\cite{sampaio2018elastic}.\\
In this section we present the geometrical properties of the EB of the FK clusters of the Ising model. Let us define the Ising model on the $L\times L$ tilted square lattice. We impose open/periodic boundary conditions along vertical/horizontal directions respectively. Then by Monte Carlo simulations of the Ising model at $h=0$, we generated $10^5$ Ising configurations at temperatures $T\leq T_c$ for $L=362,512,724,1024,1448$ and $2048$. After identifying the FK clusters, the backbones and the elastic backbones are extracted using the burning algorithm~\cite{herrmann1984backbone}. The statistics of the density of these clusters as well as the loops in the backbone are calculated. Various fractal dimensions of EBs as functions of temperature are obtained. \\
Our main observation is that there is a temperature, namely $T_{\text{eb}}<T_c$, at which the EBs undergo a phase transition from the dilute phase to the dense one. Below this temperature the elastic backbones are dense. At this temperature, the density of the EB exhibits strong large fluctuations, which signals a second order phase transition. Based on these observations we propose that there are three regimes in the zero magnetic field Ising model: for $T>T_c$ there is no spanning cluster, whereas for $T_{\text{eb}}<T<T_c$ we are in percolation regime with dilute EBs, and for $T\leq T_{\text{eb}}$ the EBs become dense. At $T=T_{\text{eb}}$ the system shows critical behavior with some critical exponents which are extracted analyzing the scaling relations.\\

\begin{figure*}
	\centering
	\begin{subfigure}{0.49\textwidth}\includegraphics[width=\textwidth]{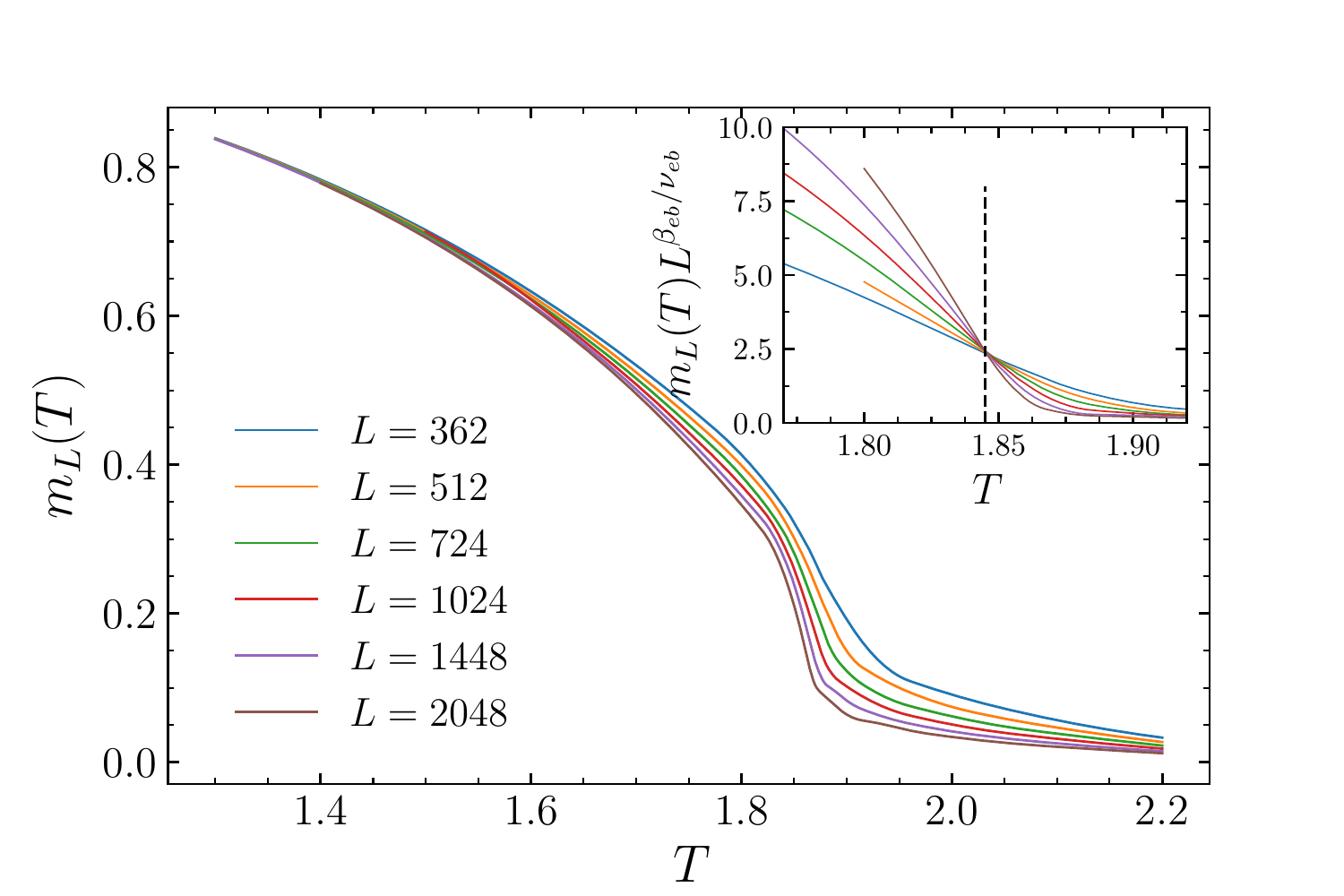}
		\caption{}
		\label{fig:Ml}
	\end{subfigure}
	\begin{subfigure}{0.49\textwidth}\includegraphics[width=\textwidth]{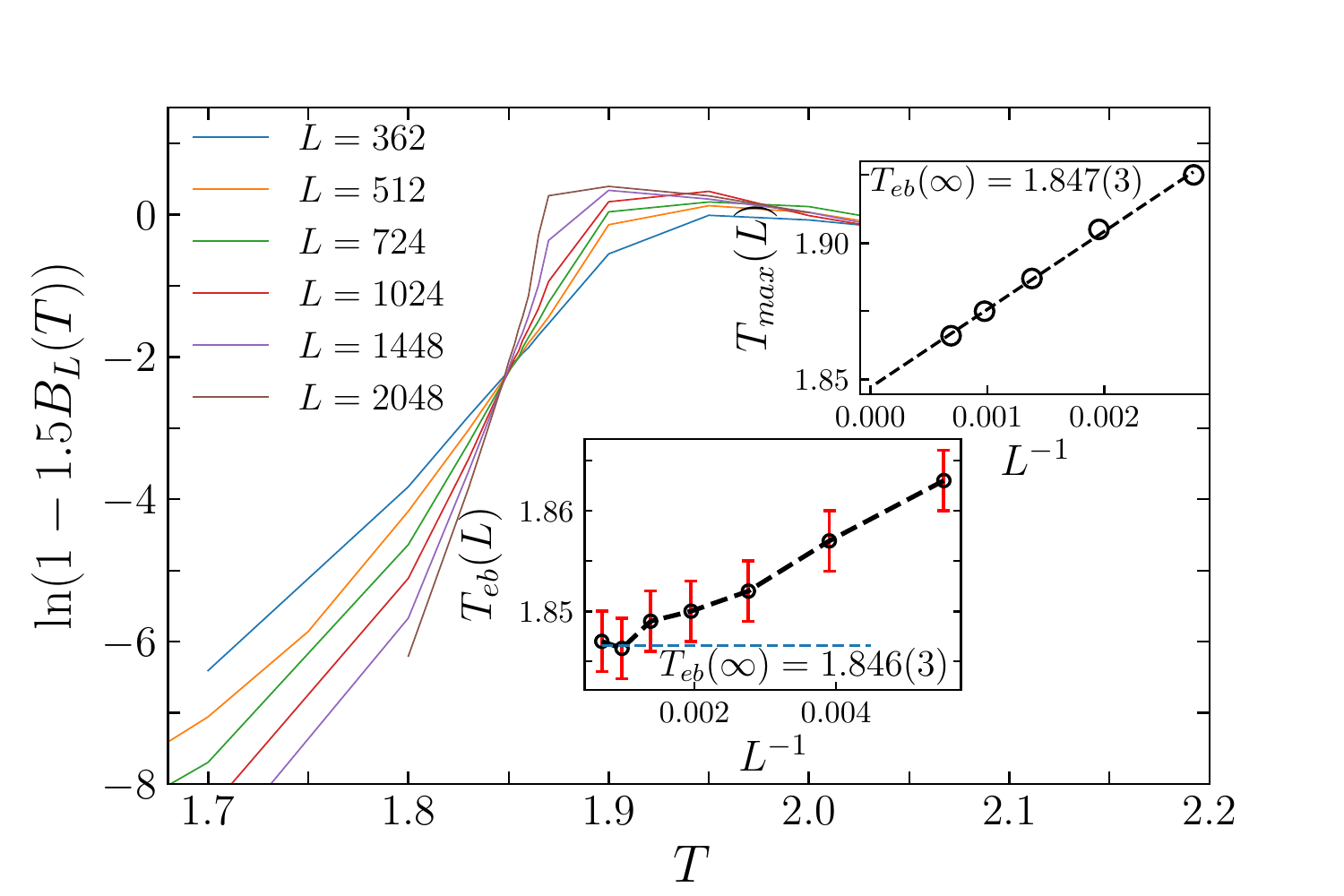}
		\caption{}
		\label{fig:Ul}
	\end{subfigure}
	\caption{(Color online): (a) The density of the elastic backbone $m_L$ in terms of temperature $T$ for various lattice sizes $L$. Inset: $L^{\beta_{\text{eb}}/\nu_{\text{eb}}}m_L(T)$ in terms of $T$ showing that $T_{\text{eb}}=1.845\pm 0.003$ and $\frac{\beta_{\text{eb}}}{\nu_{\text{eb}}}=0.48\pm 0.03$. (b) Binder's cumulant in terms of temperature $T$. $T_{\text{eb}}(L)$ is obtained as the point in which two successive graphs (for subsequent sizes) cross. This analysis shows that (lower inset) $T_{\text{eb}}=1.846\pm 0.003$. In the upper inset we show the point at which $B_L$ attains its maximum, which extrapolates to $T_{\text{eb}}$.}
	\label{fig:Teb}
\end{figure*}

In Fig.~\ref{Fig:IsingSamples} we show samples of EBs (blue) on in the tilted square lattice for three cases: at $T=T_c$ (dilute phase)~\ref{fig:HighT}, at $T=T_{\text{eb}}$ (critical value)~\ref{fig:T_eb} and at $T=1.70 <T_{\text{eb}}$ (dense phase)~\ref{fig:LowT}. The blue traces are simply the shortest paths from top to bottom. Periodic boundary conditions have been imposed in horizontal direction.\\
The first quantity to be investigated is $m_L(T)\equiv L^{-2}\left\langle M\right\rangle $ in which $M$ is the number of sites contained in the EB, and $\left\langle \right\rangle $ is the ensemble average. We consider it as the order parameter in this problem. Figure~\ref{fig:Ml} shows $m_L$ in terms of $T$ for various sizes $L$, exhibiting a clear transition at some temperature, below which the EBs become dense. By tracking the behavior of $m_L$ in terms of $T$ and $L$, one can extract the critical temperature $T_{\text{eb}}$, as done in the inset. From this analysis we observe that $T_{\text{eb}}=1.847\pm 0.001$. Also one can obtain the exponent $\beta_{\text{eb}}/\nu_{\text{eb}}$ which is obtained to be $0.48\pm 0.03$ through the scaling relation:
\begin{equation}
m_L(\epsilon)=L^{-\beta_{\text{eb}}/\nu_{\text{eb}}}G_m(\epsilon L^{1/\nu_{\text{eb}}}),
\label{Eq:scaling1}
\end{equation}
in which $G_m(x)$ is a scaling function with $G_m(x)|_{x\rightarrow \infty}\propto x^{\beta_{\text{eb}}}$ and is analytic and finite as $x\rightarrow 0$ (or equivalently $T\rightarrow T_{\text{eb}}$), and $\epsilon\equiv \frac{T_{\text{eb}}-T}{T_{\text{eb}}}$. We note here that since the system is anisotropic, one should calculate $\nu_{||}$ (the exponent parallel to the time direction) and $\nu_{\perp}$ (perpendicular exponent) separately. The relation between these anisotropic exponent and $\nu_{\text{eb}}$ will be studied at the end of this section.\\

\begin{figure}
	\centerline{\includegraphics[scale=.55]{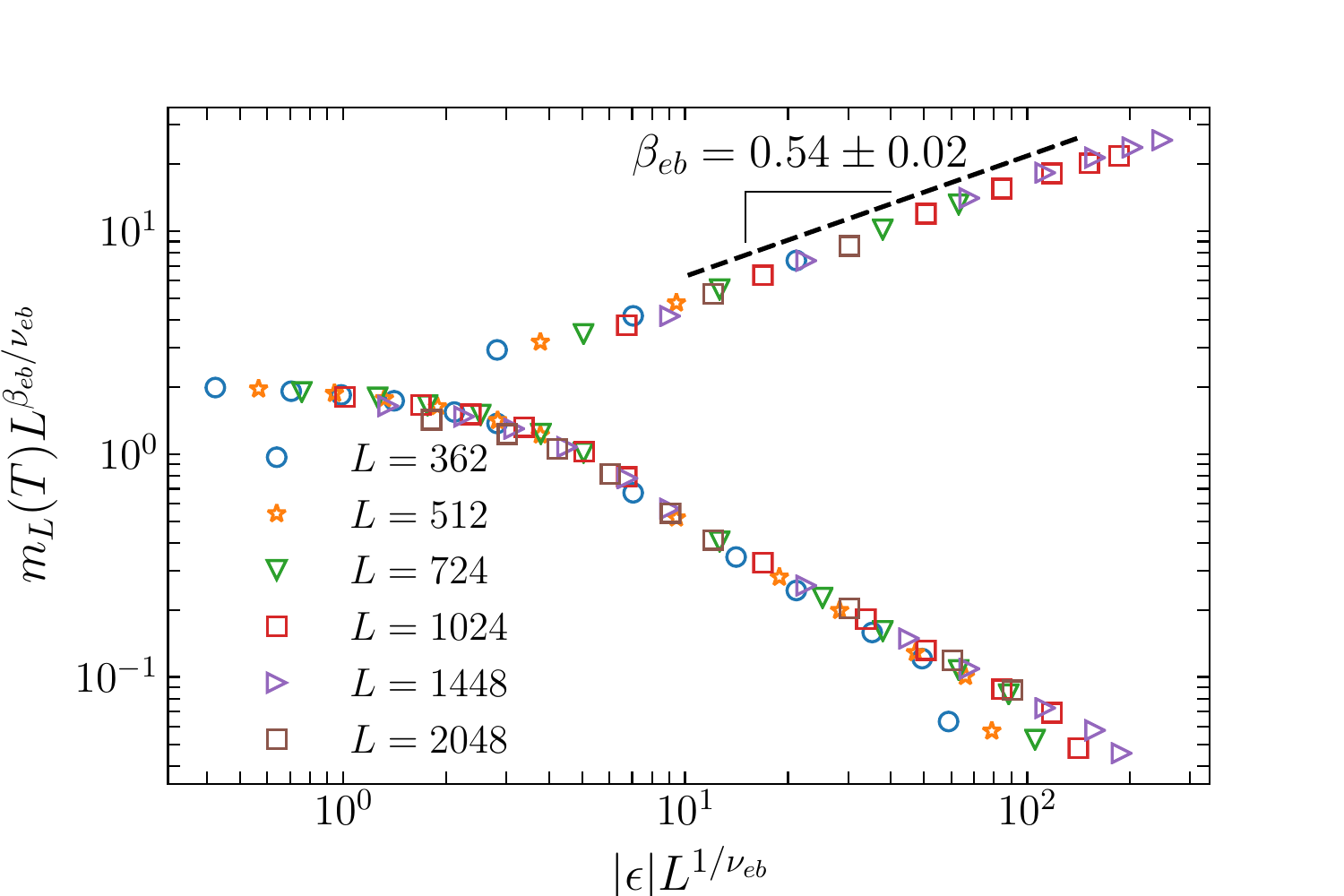}}
	\caption{(Color online): The data collapse 
		 for $m_L$. The upper branch is for $T<T_{\text{eb}}$, and the lower branch is for $T>T_{\text{eb}}$, showing that $\beta=0.54\pm 0.02$.}
	\label{fig:beta}
\end{figure}

\begin{figure}
	\centerline{\includegraphics[scale=.55]{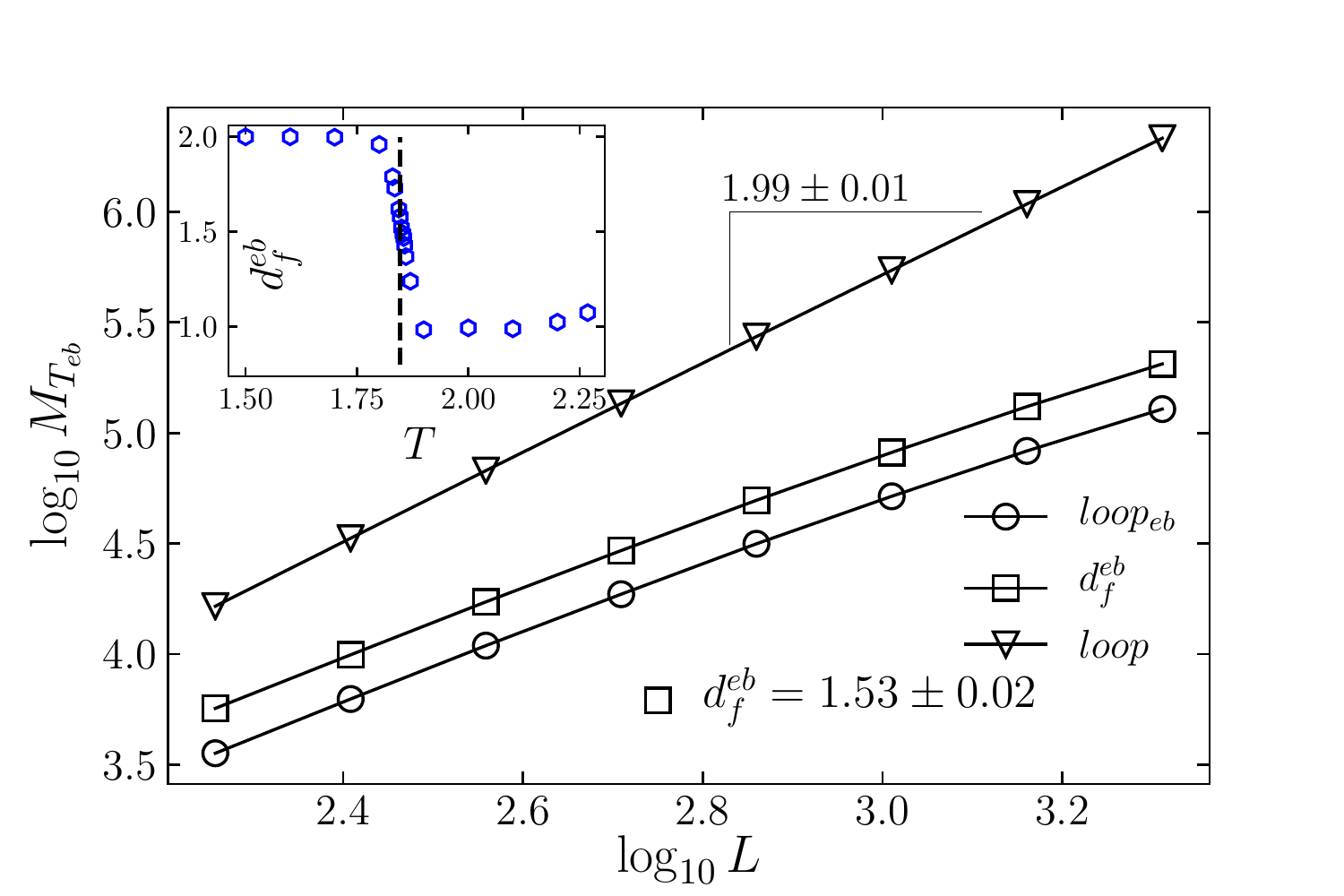}}
	\caption{(Color online): The log-log plot of $M_{\text{eb}}$ in terms of $L$. The square symbols are for the mass of the EBs, whereas circles are for the number of loops inside the EBs and the inverse triangles represent the number of loops inside the backbones. Inset: $d_f$ in terms of temperature $T$, showing that $d_f$ is within numerical accuracy $2$ for $T<T_{\text{eb}}$, and $1$ for $T>T_{\text{eb}}$.}
	\label{fig:df}
\end{figure}

To extract $T_{\text{eb}}$, one may need a more precise method. We have used the Binder's cumulant:
\begin{equation}
B_L=1-\frac{\left\langle m_L^4\right\rangle }{3\left\langle m_L^2\right\rangle^2 }
\end{equation}
which becomes $L$-independent at $T=T_{\text{eb}}$. In fact, the crossing point of two successive sizes $L$'s may change as $L$ increases, i.e. the crossing points are $L$-dependent. In this case one can extrapolate the $T_{\text{eb}}(L)$ to find the correct value, i.e. $T_{\text{eb}}(\infty)$ which is done in the lower inset of Fig.~\ref{fig:Ul}. This analysis confirms the finding of Fig.~\ref{fig:Ml}, i.e. reveals that $T_{\text{eb}}=1.846\pm 0.003$. \\
An important test is to examine whether the scaling relation Eq.~\ref{Eq:scaling1} holds or not, which is necessary for a second order transitions. We plot $G_m(x\equiv |\epsilon|L^{1/\nu_{\text{eb}}})=L^{\beta_{\text{eb}}/\nu_{\text{eb}}}$ for $T>T_{\text{eb}}$ and $T\leq T_{\text{eb}}$ to extract the exponents. This analysis has been done in Fig.~\ref{fig:beta} in which the upper branch is for $T\leq T_{\text{eb}}$, and the lower branch for $T> T_{\text{eb}}$. It is seen that for large enough $x$'s (for which we expect $G(x)\propto x^{\beta_{\text{eb}}}$), the slope is $\beta_{\text{eb}}=0.54\pm 0.02$, and also that $\nu_{\text{eb}}$ is $1.20\pm 0.03$. This implies that $\beta_{\text{eb}}/\nu_{\text{eb}}=0.47\pm 0.03$ which is compatible with the value found above.\\ 
\\
The total mass of the EBs is expected to behave like $M_{\text{eb}}=L^2m_{\text{eb}}=L^{d_f}G_m(\epsilon L^{1/\nu_{\text{eb}}})$ in which $d_f=2-\beta_{\text{eb}}/\nu_{\text{eb}}$, and $G_m$ is the same function as Eq.~\ref{Eq:scaling1}. $d_f$ is therefore obtained by a log-log plot of $M_{\text{eb}}$ in terms of $L$ which has been done for $T=T_{\text{eb}}$ in Fig.~\ref{fig:df}. We have additionally plotted the same graphs for the \textit{number of loops} inside the elastic backbone (circles) and backbones (inverse triangles). In the burning algorithm a loop is identified each time when a site is simultaneously burned from two sites and the number of loops involving $i$ and $j$ is $n-1$ when there are $n$ distinct paths from $i$ to $j$~\cite{herrmann1984backbone,herrmann1984building}. The resulting fractal dimensions are $d_f^{(1)}=1.53\pm 0.03$ and $d_f^{(2)}=1.99\pm 0.01$ respectively. We see that interestingly the fractal dimension for the number of loops inside the EBs is the same as $d_f$, and the number of loops inside the backbone grows extensively, showing that the backbone is in the dense phase. Actually we expect this for all $T<T_c$, since the backbones behave like the total FK clusters which are in dense phase in this regime. The analysis of the fractal dimension for the other temperatures shows that $d_f\approx 2.0$ for $T<T_{\text{eb}}$ (dense phase), and $d_f\approx 1.0$ for $T>T_{\text{eb}}$ (dilute phase), see the inset of Fig.~\ref{fig:df}. This confirms that the clusters are space filling for the first case, and effectively one-dimensional in the dilute phase.\\

\begin{figure*}
	\centering
	\begin{subfigure}{0.49\textwidth}\includegraphics[width=\textwidth]{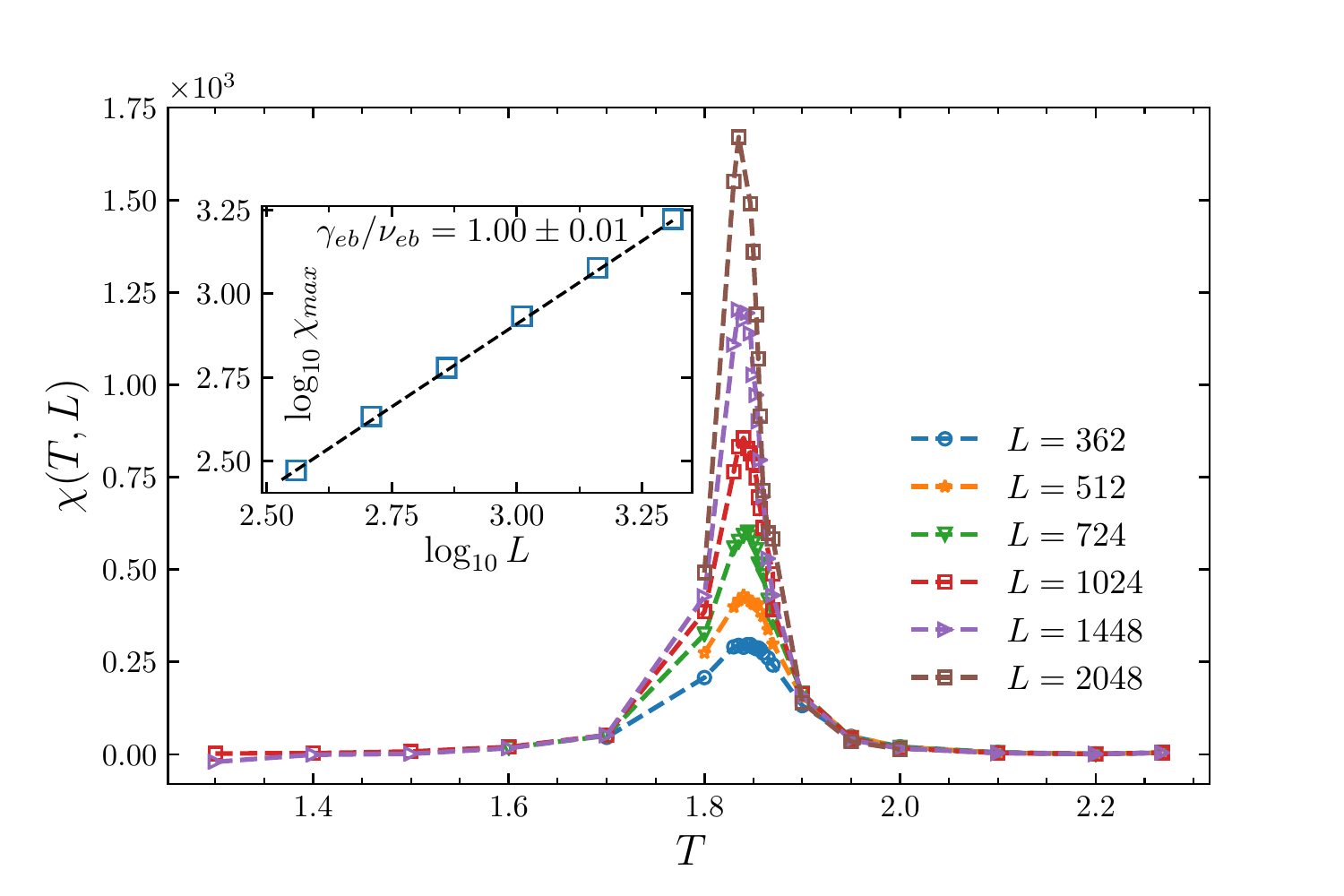}
		\caption{}
		\label{fig:gamma2}
	\end{subfigure}
	\begin{subfigure}{0.49\textwidth}\includegraphics[width=\textwidth]{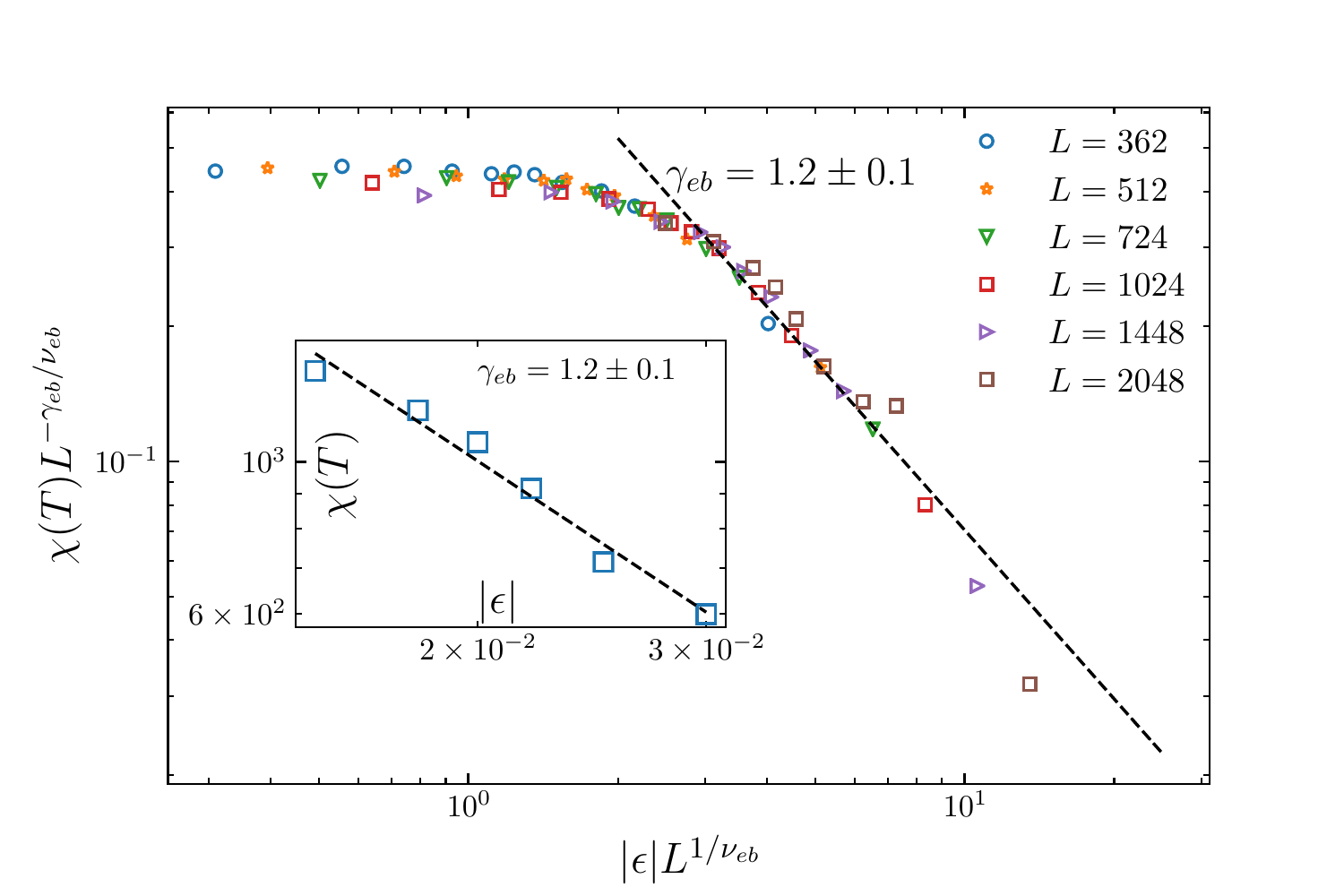}
		\caption{}
		\label{fig:gamma}
	\end{subfigure}
	\caption{(Color online): (a) $\chi(T,L)$ in terms of $T$ around $T_{\text{eb}}$ for various system sizes $L$. Inset $\chi_{\text{max}}$ (the maximum value of $\chi$ that occurs at $T_{\text{eb}}(L)$). (b) The data collapse for $\chi$ showing that $\gamma_{\text{eb}}=1.2\pm 0.1$. Inset: log-log plot of $\chi(T)$ in terms of $|\epsilon|\equiv|(T-T_{\text{eb}})/T_{\text{eb}}|$ for the largest $L$ value, i.e. $L=2048$, giving the exponent $\gamma_{\text{eb}}=1.2\pm 0.1$, confirming the data collapse analysis.}
	\label{fig:gammaha}
\end{figure*}

Given the above data, the question arises concerning the presumable singular behavior of the fluctuations of the order parameter, $m_L$, as for any second order phase transition. Let us define the fluctuation of the order parameter $\chi\equiv L^2 (\left\langle M_{\text{eb}}^2\right\rangle-\left\langle M_{\text{eb}}\right\rangle^2) $, which is expected to diverge at the transition point of any continuous transition. It is additionally expected to fulfill the scaling behavior:
\begin{equation}
\chi_L(\epsilon)=L^{-\gamma_{\text{eb}}/\nu_{\text{eb}}}G_{\chi}(\epsilon L^{1/\nu_{\text{eb}}}),
\label{Eq:scaling2}
\end{equation}
in which again $G_{\chi}(x)$ is a scaling function with $G_{\chi}(x)|_{x\rightarrow \infty}\propto x^{\gamma_{\text{eb}}}$ and is analytic and finite as $x\rightarrow 0$. The analysis of this function is presented in Fig.~\ref{fig:gammaha}. This scaling hypothesis predicts that the maximum value of $\chi$, i.e. at the transition point behaves like $\chi_{\text{max}}\propto L^{\gamma_{\text{eb}}/\nu_{\text{eb}}}$, and also $\chi_L(\epsilon)\propto \left| \epsilon\right|^{-\gamma_{\text{eb}}}$ for small enough $|\epsilon|$. Fig.~\ref{fig:gamma2} suggests that $\gamma_{\text{eb}}/\nu_{\text{eb}}=1.00\pm 0.01$. If we use the above-obtained $\nu_{\text{eb}}$ ($1.20\pm 0.03$), we find that $\gamma_{\text{eb}}=1.20\pm 0.03$. Summarizing we have presented the data collapse analysis in Fig.~\ref{fig:gamma} which confirms that $\gamma_{\text{eb}}=1.2\pm 0.1$. The inset is also consistent with this result. \\

Here it is worthy to comment on the hyper-scaling relations. As mentioned in SEC.~\ref{FKCluster} the exponents are not independent, and there are some hyper-scaling relations between them. For example, $d_f=2-\beta/\nu$, and $2\beta+\gamma=d\nu$ ($d=2$ here). The latter has been shown to be violated for the EB transition in percolation~\cite{sampaio2018elastic}. Here we note that $\beta_{\text{eb}}+\frac{1}{2}\gamma_{\text{eb}}=1.15 \pm 0.07 $ which agrees within the error bar with $\nu_{\text{eb}}=1.2 \pm 0.1$. Therefore, we conclude that the hyper-scaling relation is restored in the FK clusters of Ising model.\\
 
The set of all shortest paths leaving one point can be seen as an anisotropic object, and the corresponding critical point (the transition point) should be described by an anisotropic universality class. Recently it was suggested by Deng \textit{et al.}~\cite{deng2018elastic} that the transition point of the EBs defined in the percolation system is in the universality class of DP. To this end, they calculated two fractal dimensions for both the EB of percolation and the backbone of DP: firstly the number of occupied sites along the center line $N_b\equiv \left\langle N_{y=L/2}\right\rangle$ (which represents the behavior of the bulk) and the number of occupied sites at the top and bottom edges $N_e\equiv \frac{1}{2}\left\langle N_{y=1}+N_{y=L}\right\rangle $ (representing the behavior of boundaries) in terms of system size $L$, and secondly the chemical distance (shortest path) exponent $d_{\text{min}}$ defined by $\left\langle l_s\right\rangle \sim L^{d_{\text{min}}}$. From the similarities between the obtained fractal dimensions and the exponents of the DP ($d_{DP}=2-\frac{\beta}{\nu_{||}}$ characterizing the full DP, and $d_{B,DP}=2-\frac{\beta}{\nu_{||}}-\delta$ characterizing the bulk of the DP, in which the exponent $\delta$ is defined by the survival probability $P(t)\propto t^{-\delta}$), Deng \textit{et al.} concluded that they are in the same universality classes. Note that $N_e$ and $N_b$ are expected to scale like $L^{d_e-1}$ and $L^{d_b-1}$ (the subtraction of exponents by one is due to the fact that we are taking one-dimensional cuts through the clusters). The above described procedure still requires some consistent derivation, e.g. anisotropic scaling should be tested. However we do the same analysis here to calculate $d_e$ and $d_b$ as in Ref.~\cite{deng2018elastic}.\\
\begin{table*}
	\begin{tabular}{c | c c c c c c c c c c }
		\hline exponent & $\beta$ & $\nu=\nu_{\perp}$ & $\nu_{||}$ & $\gamma$ & $2-\beta/\nu$ & $d_f$ & $d_e$ & $d_b$ & $d_{\text{min}}$ & $\frac{\beta}{\nu}+\frac{\gamma}{2\nu}$ \\
		\hline Fig.~\ref{fig:Ml}a,~\ref{fig:df},~\ref{fig:fractalDimensions} & -- & -- & -- & -- & $1.52(3)$ & $1.53(2)$ & $1.71(1)$ & $1.52(1)$ & $1.090(4)$ & -- \\
		\hline Fig.~\ref{fig:beta} & $0.54(2)$ & $1.20(3)$ & -- & -- & $1.53(3)$ & -- & -- & -- & -- & -- \\
		\hline Fig.~\ref{fig:gamma2},~\ref{fig:N_eb} & -- & $1.21(4)$  & $1.86(1)$ & $1.20(3)$ & -- & -- & -- & -- & -- & $0.95(3)$ \\
		\hline OP ($p=p_{\text{eb}}$)
		& $0.50(2)$ & $2.00(2)$  & -- & $1.97(5)$ & $1.750(3)$ & $1.750(3)$ & $1.84054(4)$ & $1.68102(15)$ & -- & $0.74(1)$ \\
		\hline OP ($p=p_c$)
		& $\frac{5}{36}\approx 0.14$ & $\frac{4}{3}\approx 1.33$  & -- & $\frac{43}{18}\approx 2.39$ & $d_f$ & $\frac{91}{48}\approx  1.896$ & -- & -- & $1.13077(2)$~\cite{herrmann1988fractal,grassberger1992spreading,dokholyan1998scaling,newman2000efficient} & $1$ \\
		\hline DP ($p=p_c$)
		& $0.277(2)$ & $1.0969(3)$  & $1.7339(3)$ & -- & $1.747(3)$ & $1.765(1)$ & -- & -- & -- & -- \\
		\hline
	\end{tabular}
	\caption{The exponents for the Ising model at $T=T_{\text{eb}}$ (rows: $2, 3$ and $4$), for ordinary percolation (OP) model at $p=p_{\text{eb}}$ (row $5$)~\cite{sampaio2018elastic}, for OP at $p=p_c$ (row $6$, in which the exact results can be found in~\cite{den1979relation,pearson1980conjecture,nienhuis1980magnetic,nienhuis1984coulomb,cardy1984conformal,saleur1987exact,grossman1987accessible,cardy1998number}, and are numerically confirmed in~\cite{sykes1974percolation,nakanishi1980scaling,levinshteln1975relation}), and finally for DP model at $p=p_c$~\cite{hede1991self}. For OP$_{p_{\text{eb}}}$ and DP$_{p_c}$, although the sole values of $\beta$ and $\nu$ are different, $\beta/\nu$ (which is equal to $\beta/\nu_{\perp}$) is the same. The exponents for the Ising model are considerably different from the exponents of DP, therefore define a new anisotropic universality class. Two hyper-scaling relations ($d_f=2-\beta/\nu$ and $2\beta+\gamma=2\nu$) are also reported in the table, which are shown to be valid for the Ising model, whereas the latter is violated for OP$_{p_{\text{eb}}}$.}
	\label{tab:exponents}
\end{table*}

\begin{figure*}
	\centering
	\begin{subfigure}{0.49\textwidth}\includegraphics[width=\textwidth]{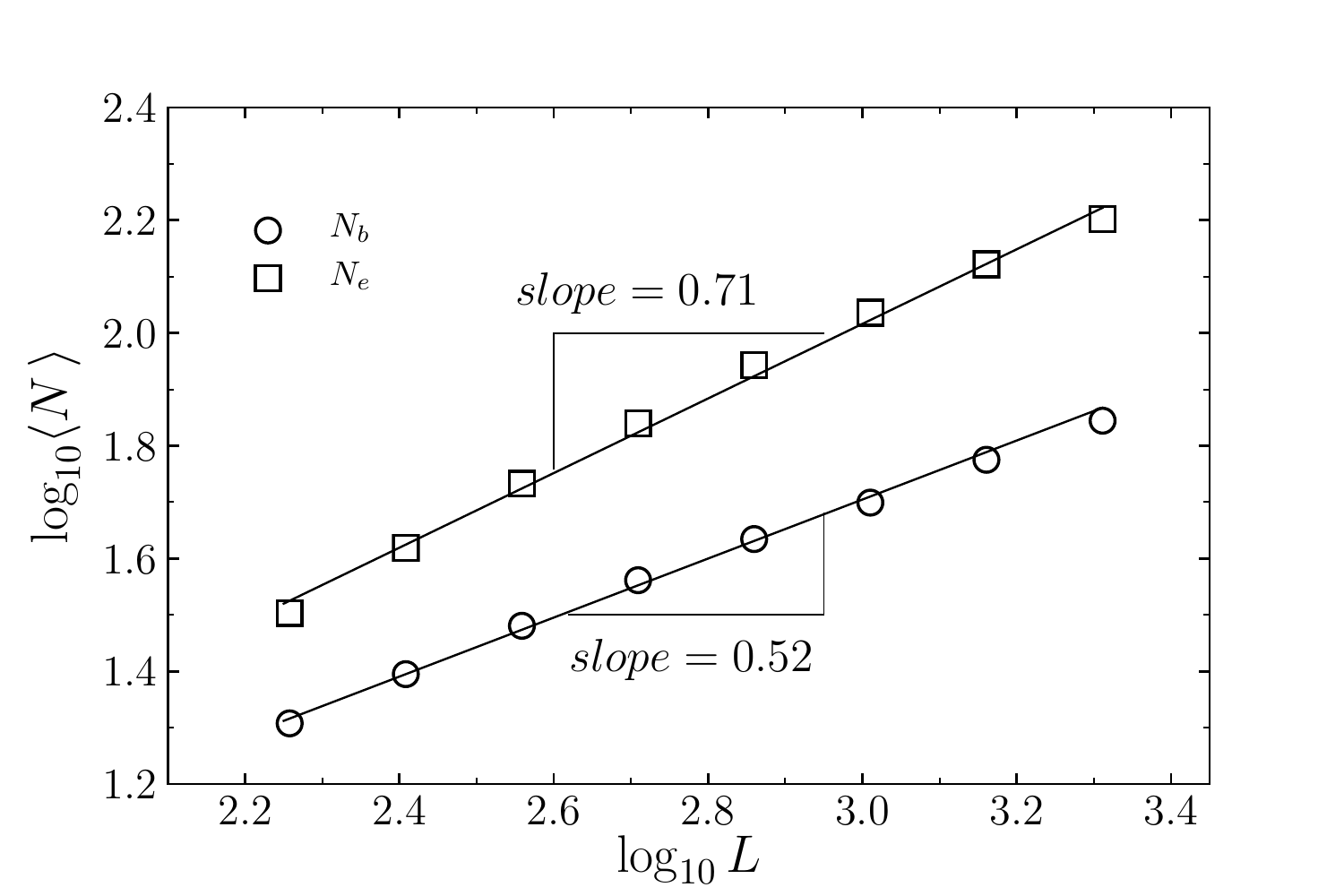}
		\caption{}
		\label{fig:N_eb}
	\end{subfigure}
	\begin{subfigure}{0.49\textwidth}\includegraphics[width=\textwidth]{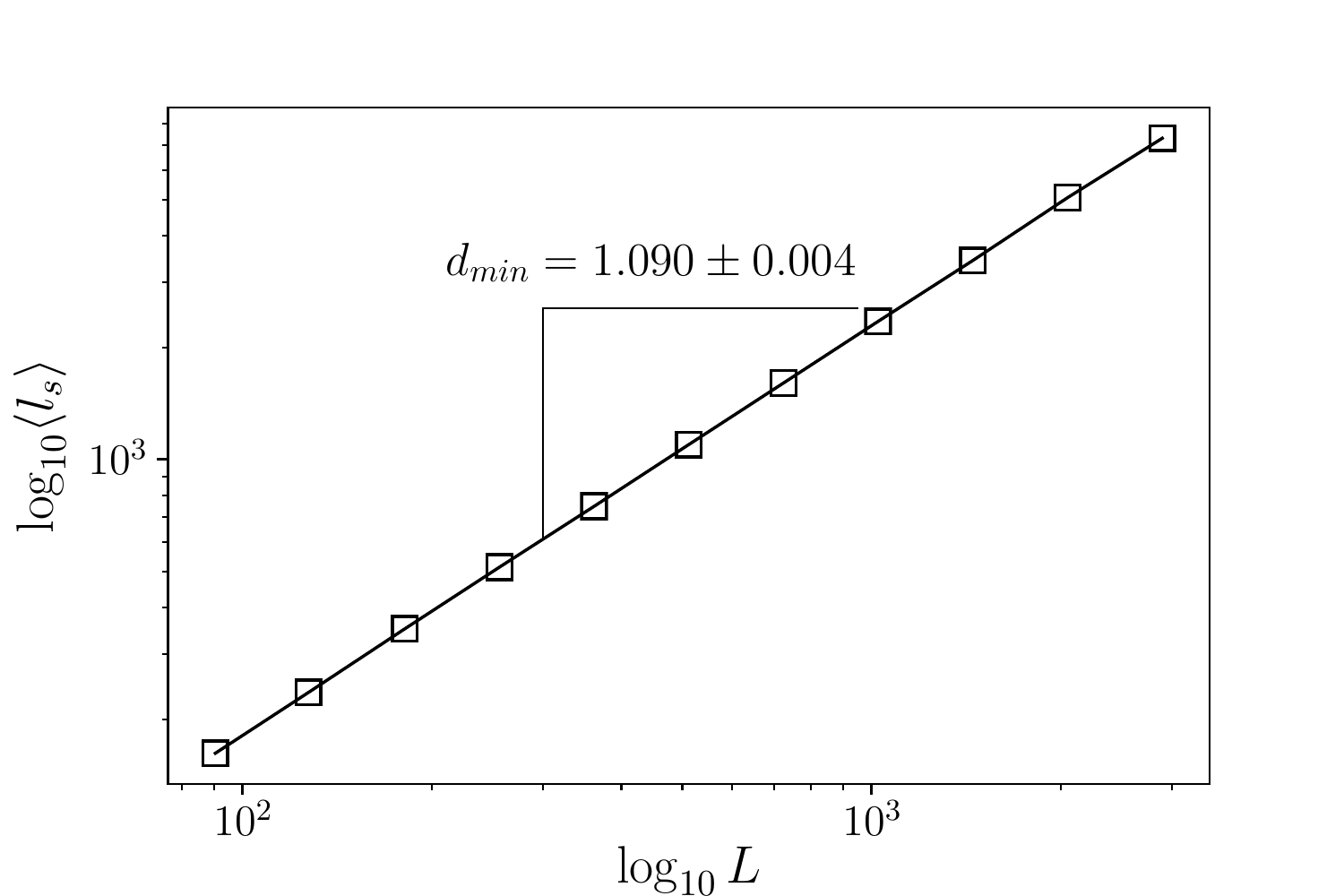}
		\caption{}
		\label{fig:shortespath_Df}
	\end{subfigure}
	\caption{(a) Log-log plot of $N_b$ and $N_e$ in terms of system size $L$ giving exponents $d_e=1.71\pm0.01$ and $d_b=1.52\pm 0.01$. (b) The fractal dimension corresponding to the shortest path $d_{\text{min}}=1.090\pm 0.004$. }
	\label{fig:fractalDimensions}
\end{figure*}

Such an analysis at $T=T_{\text{eb}}$ shows that the universality class is very different from DP, and belongs to another anisotropic universality class. From the Figs.~\ref{fig:N_eb} and~\ref{fig:shortespath_Df} we conclude that $d_e=1.71\pm 0.01$ and $d_b=1.52\pm 0.01$. Therefore $\frac{\beta}{\nu_{||}}=0.29\pm 0.01$ resulting in $\nu_{||}=1.86 \pm 0.01$. Additionally, if the reasoning of equations of the DP exponents is applicable here, then the exponent of the survival probability will be $\delta=0.19\pm 0.01$. There are some proposals concerning the relation between $d_f$ and $\nu_{||}$ and $\nu_{\perp}$ for DP~\cite{hede1991self}. If one uses the most accepted one, i.e. $d_f=2-\beta/\nu_{\perp}$~\cite{hede1991self}, then it results in $\nu_{\perp}=\nu=1.21\pm 0.04$, which is compatible with the general expectation that the ratio of correlation lengths vanishes in the thermodynamic limit, i.e. $\zeta_{\perp}/\zeta_{||}\rightarrow 0$ when $L\rightarrow \infty$. \\

All exponents are presented in TABLE~\ref{tab:exponents}. For comparison, the same exponents are shown for $p=p_c$ and $p=p_{\text{eb}}$. Although the $\beta$ exponent for Ising model and percolation ($p=p_{\text{eb}}$) are close to each other, the other exponents are drastically different. The other exponent that is relevant in characterizing the geometrical properties of the model at $T=T_c$ is the fractal dimension of the shortest path ($d_{\text{min}}$). This dependence is shown in Fig.~\ref{fig:shortespath_Df}, from which we see that $d_{\text{min}}=1.090\pm 0.004$. This exponent has perviously been conjectured by Deng \textit{et. al.} to be $1.09375$~\cite{deng2010some} and numerically calculated by Hou \textit{et. al.} where the value $1.0940(3)$ was reported~\cite{hou2019geometric}. This value should also be compared with $d_{\text{min}}^{\text{percolation}}(p=p_c)$ which is $1.13077(2)$~\cite{herrmann1988fractal,grassberger1992spreading,dokholyan1998scaling,newman2000efficient} , i.e. the shortest paths are less tortuous for the FK clusters of Ising model.\\

One may be interested in calculating $\sigma$ and $\tau$. We obtain $\tau=(2d\nu-\beta)/(d\nu-\beta)=2.29\pm 0.02$, and $\sigma=1/(d\nu-\beta)=0.53\pm 0.02$. Since our model is anisotropic it will not be conformally invariant~\cite{Bauer2003Conformal} and a Loewner transformation
would map its paths to anomalous diffusion~\cite{credidio2016stochastic}.

\section*{Discussion and Conclusion}
\label{sec:conc}
 The elastic backbone of the Ising model (in the zero magnetic field limit) has numerically been considered in this work. The geometrical properties of the critical models are coded in the FK clusters, which are obtained simply by dilution of the geometrical clusters of same spin. {Based on our numerical evidences we proposed} that the elastic backbone of the FK clusters undergoes a continuous transition at some temperature $T_{\text{eb}}<T_c$. $m_L\equiv L^{-2}M_L$ (being the average number of sites of the elastic backbone of the spanning FK clusters in a system of linear size $L$) has been considered as the order parameter for this transition. Using Binder's cumulant we found $T_{\text{eb}}=1.846\pm 0.003$. We have obtained $\beta$ and $\nu$ exponents using various methods, which yield consistent values. The exponents are different from both critical percolation, and the percolation at $p=p_{\text{eb}}$, i.e. $2-\beta_{\text{eb}}/\nu_{\text{eb}}=1.52\pm 0.03$. The determination of other exponents (for example $\gamma$ obtained from the density fluctuations, $d_f,d_e,d_b$ and $d_{\text{min}}$) reveals that the universality class of this transition is considerably different from ordinary percolation at $p=p_c$ and $p=p_{\text{eb}}$, and also the Ising model at $T=T_c$. We have characterized comprehensively exponents which seem to be in a new universality class for anisotropic systems. The parallel correlation length exponent $\nu_{||}$ and $d_{\text{min}}$ were found to be $1.86\pm 0.01$ and $1.090\pm 0.004$ respectively which are different from the ones for DP ($1.7339(3)$ and $1.13077(2)$ respectively). Importantly we have shown that two relevant hyper-scaling relations hold here, one of which is violated for percolation at $p=p_{\text{eb}}$.

\bibliography{refs}

\end{document}